\documentclass[lettersize,journal]{IEEEtran}
\ifCLASSINFOpdf

\else
\fi
\usepackage{url}
\usepackage{comment}
\includecomment{toexclude}
\usepackage{amsmath,amsfonts,amssymb}
\usepackage{cite}
\usepackage{verbatim}
\usepackage{bm}
\usepackage{graphicx}
\usepackage{xcolor}
\usepackage{subfigure}
\usepackage{diagbox}
\usepackage{multirow}
\usepackage{array}
\usepackage{statmath}
\usepackage[belowskip=-5pt,aboveskip=0pt]{caption}
\newcolumntype{P}[1]{>{\centering\arraybackslash}p{#1}}
\usepackage{flushend}

\usepackage{enumitem}
\usepackage{epstopdf}
\usepackage{algorithm,algpseudocode}
\usepackage[justification=centering]{caption}

\usepackage{xcolor}
\graphicspath{{figures/}}

\usepackage{mathtools}

\algrenewcommand\algorithmicindent{0.7em}%
\DeclareMathAlphabet\mathbfcal{OMS}{cmsy}{b}{n}
\usepackage{stackengine}
\def\delequal{\mathrel{\ensurestackMath{\stackon[1pt]{=}{\scriptstyle\Delta}}}}
\begin{document}
	\title{Frequency Hopping Joint Radar-Communications with Hybrid Sub-pulse Frequency and Duration Modulation}
	\author{Linh~Manh~Hoang,~\IEEEmembership{Student Member,~IEEE}, J.~Andrew~Zhang,~\IEEEmembership{Senior Member,~IEEE}, Diep~N.~Nguyen,~\IEEEmembership{Senior Member,~IEEE}, and Dinh Thai Hoang,~\IEEEmembership{Senior Member,~IEEE}
	\thanks{The authors are with the School of Electrical and Data Engineering, University of Technology Sydney, NSW 2007. e-mail: Linh.M.Hoang@student.uts.edu.au; (Andrew.Zhang, Diep. Nguyen, Hoang.Dinh)@uts.edu.au}}

	\maketitle
	\begin{abstract}
Frequency-hopping (FH) joint radar-communications (JRC) can offer excellent security for integrated sensing and communication systems. However, existing JRC schemes mainly embed information using only the sub-pulse frequencies and hence the data rate is limited. In this paper, we propose to use both sub-pulse frequencies and durations for information modulation, leading to higher communication data rates. For information demodulation, we propose a novel scheme by using the time-frequency analysis (TFA) technique and a `you only look once' (YOLO)-based detection system. As such, our system does not require channel estimation, simplifying the transmission signal frame design. Simulation results demonstrate the effectiveness of our scheme, and show that it is robust against the Doppler shift and timing offset between the transceiver and the communication receiver.
\end{abstract}
	\IEEEpeerreviewmaketitle
	\vspace{-0.3cm}
	\section{Introduction}
The frequency-hopping (FH) joint radar-communications systems (JRC)  \cite{hassanien2019dual,zhang2021overview,huang2020majorcom,wu2020waveform,baxter2018dual} have become more and more prevalent in both military and civil applications, e.g., airborne, shipborne, ground-based combat systems, the connected autonomous vehicle (CAV). The most popular FH JRC systems include multiple-input multiple-output orthogonal frequency division multiplexing (MIMO-OFDM), multi-Carrier AgilE phaSed Array Radar (CAESAR), and FH-MIMO\cite{zhang2021overview}.

    
    A key challenge in FH JRC is to embed/modulate data (onto frequency sub-pulses) at the transceiver and demodulate the signal at the receiver. For data embedding, the most dominant technique is the index modulation (IM), which utilizes frequency selection/combination and/or antenna selection/permutation for data representation. Specifically, the data bits are embedded by selecting different sets of sub-pulse frequency (i.e., frequency combination) and allocating them to different antennas (i.e., antennas permutation). For data demodulation, the optimal demodulator can be based on the maximum likelihood principle \cite{huang2020majorcom}, while sub-optimal methods, which require lower computation complexities, are based on compressive sensing (CS) or discrete Fourier transform (DFT) \cite{zhang2021overview}. However, the performance of these methods depends on the accuracy of channel estimation. Unfortunately, a long training sequence necessary for accurate channel estimation is not always feasible. This is because a long training sequence reduces the communication fraction over the whole time frame, thus decreasing the data bit rate. In particular, the data embedding schemes using both sub-pulse frequency combinations and antenna permutations require complex demodulation techniques that are prone to demodulation error. On the other hand, the data embedding scheme using only the sub-pulse frequency combination has a limited data transmission rate.
    
    This paper proposes novel techniques to embed and demodulate data in an FH JRC system to increase the data rate and reduce the demodulation error. For data embedding, we use both sub-pulse frequency and duration, therefore increasing the data transmission rate compared to only using the sub-pulse frequency. For data demodulation, we propose a novel scheme based on the signal's time-frequency image (TFI) and a YOLO-based detection system. This demodulation scheme does not require channel estimation and is robust to the Doppler shift and the timing offset between the transceiver and the communication receiver. Moreover, the proposed data embedding and demodulation schemes are spatially flexible and not limited to the sidelobe of the transmit beampattern, since the data is not embedded by utilizing the phase or amplitude of the beampattern sidelobe.
    \section{System and Signals}
	\label{Sec:System_model}
	\begin{figure}[t]
		\centering
		\includegraphics[width=0.85\linewidth]{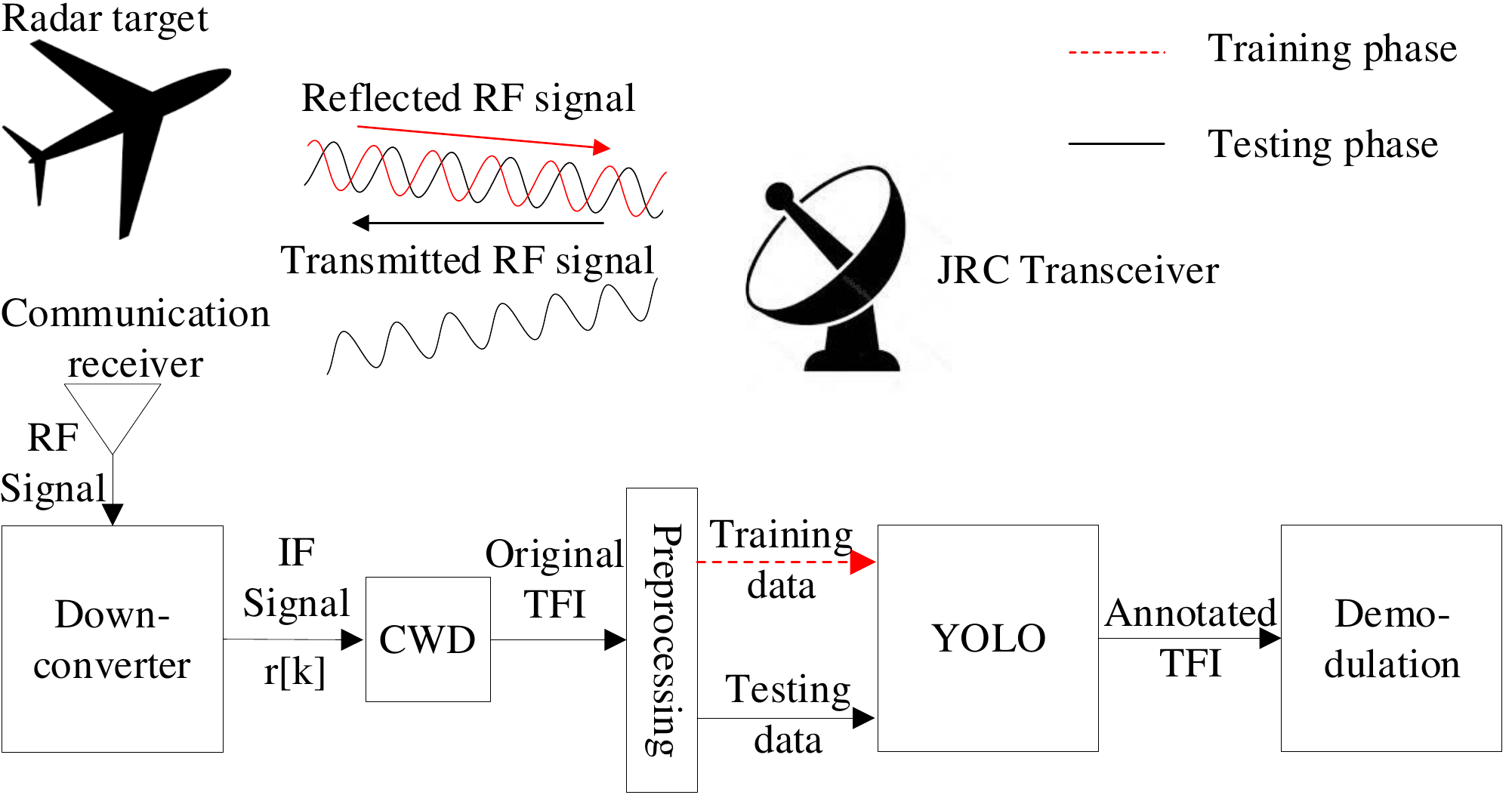}
		\vspace{0.2cm}
		\caption{FH joint radar-communications system.}
		\label{fig:block}
	\end{figure}
	\subsection{System Overview}
	We consider a ground-to-air JRC system, where the JRC transceiver is located on the ground and the communication receiver is mounted on an aircraft or a high-altitude platform (HAP), as illustrated in Fig. \ref{fig:block}. For sensing, the JRC transceiver transmits the radio frequency (RF) signal and processes the reflected RF signal to determine target's position and velocity. For data transmission, the communication receiver obtains the RF signal from the JRC transceiver and demodulate the embedded data. In such a system, the line-of-sight (LoS) path dominates, and non-LoS paths are negligible. At the transceiver or communication receiver, the received RF signal is down-converted into the intermediate frequency (IF) signal and sampled at sampling frequency $f_{\rm s}$ to generate a discrete-time signal as
	\begin{align}\label{rec_sig}
	r[k]&=s[k]+w[k].
	\end{align}
	Here, $k$ is the sample index which increases every $T_{\rm s}=1/f_{\rm s}$, $s[k]$ and $r[k]$ are the transmitted and received signal samples, respectively, and $w[k]$ is the complex additive white Gaussian noise (AWGN). The JRC system uses a pulse wave (PW) waveform with a duty cycle $D=\triangle t_{\rm w}/\triangle t_{\rm r},D\ll 1$, where $\triangle t_{\rm w}$ and $\triangle t_{\rm r}$ are the pulse width and pulse repetition intervals, respectively. The received signal at the communication receiver is then used to generate the Choi-Williams distribution time-frequency image (CWD-TFI). Next, the CWD-TFI is preprocessed in the Preprocessing block to generate the training and testing data for the YOLO detection system. The training process of the YOLO detection system is demonstrated by the dashed line in Fig. \ref{fig:block} using the training data. The trained YOLO detection system and the testing CWD-TFIs are used to generate the annotated CWD-TFIs, which are used to demodulate the embedded data.
	\vspace{-0.2cm}
	\subsection{Signals Model} 
	The JRC system uses FH signals with each signal pulse being divided into $N_{\rm f}$ sub-pulses (i.e., hops) with different frequencies and durations. Let ${\bf{F}}=(f_{1},f_{2},...,f_{N_{\rm f}})$ denote the FH sequence and ${\bf{T}}=(\triangle t_{1},\triangle t_{2},...,\triangle t_{{N_{\rm f}}})$ denote the sub-pulse duration sequence. The JRC signals in each pulse is represented by
	\begin{align}
	\label{sig_model}
	s[k]=\sum_{n=1}^{N_{\rm f}}&Ae^{j2\pi f_{n}kT_{\rm s}}{\rm {rect}}\Big[\frac{kT_{\rm s}-\sum_{i=0}^{n-1}\triangle t_i}{\triangle t_n}\Big],\\
	\text{where } A \text{ is the }& \text{signal amplitude, }\qquad\nonumber\\
	{\text {rect}}[k]&\delequal\begin{cases}
	1, \qquad 0\leq k\leq 1,\nonumber \\
	0, \qquad {\rm {otherwise}},\\
	\end{cases}\\
	\text{and } \triangle t_0&\delequal 0.\nonumber
	\end{align}
	\vspace{-0.5cm}
	\section{Data Embedding Schemes}
\label{sec:embedding}
	In this work, we use the Costas arrays to generate the FH sequence of each signal pulse in one of the data embedding schemes. Using the Costas array, an FH signal possesses a narrow peak at the origin of its ambiguity function (AF) and low sidelobes elsewhere, which is desirable for the radar measurement (i.e., range and speed) accuracy. Details of the Costas arrays, their construction, and characteristics can be found in \cite[Ch.~5]{levanon2004radar}.
	
	We propose two data embedding techniques, namely \textit{Random} and \textit{Costas-based} schemes. Unlike existing studies that only utilize sub-pulse frequencies to convey information, both sub-pulse frequencies and durations are used in our proposed schemes. We first design the codebooks $\mathcal{S}$ and $\mathcal{S}_{\rm C}$ of the \textit{Random} and \textit{Costas-based} schemes, respectively. Then, for each embedding scheme, data bits are embedded by mapping each element of the corresponding codebook to a symbol. Let ${\bf{\mathcal{F}}}\delequal[(1,2,...,N_{\rm f})\times f_{\rm f}]$ be the selection set of the sub-pulse frequency, where $f_{\rm f}$ is the fundamental frequency. Let ${\bf{\mathcal{D}}}\delequal(\triangle _{1},\triangle _{2},...,\triangle _{N_{{\rm{T}}}})$ be the selection set of the sub-pulse duration, where $N_{{\rm{T}}}$ is the number of sub-pulse duration selections. The codebooks $\mathcal{S}$ and $\mathcal{S}_{\rm C}$ are designed as follows. 
	\begin{itemize}
	\item \textit{Random} scheme: $\mathcal{S}$ is designed by selecting each sub-pulse frequency from the frequency set ${\bf{\mathcal{F}}}$, and selecting each sub-pulse duration from the duration set ${\bf{\mathcal{D}}}$ as
	\begin{align}
	f_i&\in{\bf{\mathcal{F}}},\triangle t_{i}\in{\bf{\mathcal{D}}},\forall i\in(1,2,...,N_{\rm f}),\\
	f_i&\neq f_{i+1}, \forall i\in(1,2,...,N_{\rm f}-1).\label{condition1}
	\end{align}
	Here, condition (\ref{condition1}) means every two consecutive sub-pulse frequencies are not equal, which is necessary for the data demodulation technique that will be presented in Section \ref{sec:Data Demo}.  
	\item \textit{Costas-based} scheme: Unlike the \textit{Random} scheme, for designing $\mathcal{S}_{\rm C}$, each sub-pulse frequency is not individually selected. Instead, the whole FH sequence of each signal pulse is selected from the Costas-based FH sequence defined as follows.
	
	Let ${\bf{\mathcal{A}}_{\rm C}}\delequal({\bf{A}}_1,{\bf{A}}_2,...,{\bf{A}}_{N_{\rm C}})$ be the set of Costas array of length $N_{\rm f}$, where ${\bf{A}}_i, \forall i\in (1,2,...,N_{\rm C})$ is an $\mathbb{N}^{N_{\rm f}\times 1}$ Costas array, and $N_{\rm C}$ is the number of Costas sequence with length of $N_{\rm f}$. Then, let ${\bf{\mathcal{F}}_{\rm C}}\delequal[({\bf{A}}_1,{\bf{A}}_2,...,{\bf{A}}_{N_{\rm C}})\times f_{\rm f}]$ be the selection set of the Costas-based FH sequence. Then, the FH sequence of each signal pulse is selected from ${\bf{\mathcal{F}}_{\rm C}}$. On the other hand, similar to the \textit{Random} scheme, each sub-pulse duration is selected from the sub-pulse duration set ${\bf{\mathcal{D}}}$ as
	\begin{align}
	{\bf{F}}&\in{\bf{\mathcal{F}}_{\rm C}},\triangle t_{i}\in{\bf{\mathcal{D}}},\forall i\in(1,2,...,N_{\rm f}).
	\end{align}
	\end{itemize}
	Note that for the \textit{Costas-based} scheme, because the FH sequence is generated using the Costas array, condition (\ref{condition1}) is satisfied. The dimensions of $\mathcal{S}$ and $\mathcal{S}_{\rm C}$ are given by
	\begin{align}
	|\mathcal{S}|&=\underbrace{N_{{\rm{T}}}^{N_{\rm f}}}_\text{sub-pulse duration selections}\times \underbrace{N_{\rm f}(N_{\rm f}-1)^{(N_{\rm f}-1)}}_\text{sub-pulse frequency selections},\label{random_combine}\\
	|\mathcal{S}_{\rm C}|&=\underbrace{N_{{\rm{T}}}^{N_{\rm f}}}_\text{sub-pulse duration selections}\times \underbrace{N_{\rm C}}_\text{sub-pulse frequency selections}.\label{Costas_combine}
	\end{align}
	As shown in (\ref{random_combine}), the number of selections of sub-pulse frequencies is $N_{\rm f}(N_{\rm f}-1)^{(N_{\rm f}-1)}$ instead of $N_{\rm f}^{N_{\rm f}}$ due to the requirement in (\ref{condition1}).
	
	It follows from (\ref{random_combine}) and (\ref{Costas_combine}) that the maximum number of bits $C$ and $C_{\rm C}$ that can be embedded in each pulse of the two schemes are given by
	\begin{align}
	C=&\lfloor\log_2(|\mathcal{S}|)\rfloor=\lfloor\log_2(N_{{\rm{T}}}^{N_{\rm f}}N_{\rm f}(N_{\rm f}-1)^{(N_{\rm f}-1)})\rfloor,\nonumber\\
	C_{\rm C}=&\lfloor\log_2(|\mathcal{S}_{\rm C}|)\rfloor=\lfloor\log_2(N_{{\rm{T}}}^{N_{\rm f}}N_{\rm C})\rfloor,
	\end{align}
	where $\lfloor.\rfloor$ denotes the floor function, determining the closest smaller integer. As can be seen, the \textit{Random} scheme achieves a higher transmission rate than the \textit{Costas-based} scheme.
	\vspace{-0.3cm}
	\section{Sensing at the JRC Transceiver}
This section demonstrates that by using the sub-pulse frequencies and duration for data embedding as presented in the previous section, the detection probability of the sensing function is not affected. Let $P_{\rm D}$ and $P_{\rm FA}$ denote the detection and false alarm probabilities of the JRC transceiver, respectively. Let $E_{\rm{p}}$ be the signal energy per each pulse of the JRC signal and ${\sigma_{\rm w}^2}$ be the noise variance. The energy-to-noise ratio (ENR) of the JRC signal is given by
\begin{align}
    \text{ENR}&=10\log_{10}(\frac{E_{\rm{p}}}{{\sigma_{\rm w}^2}}).
\end{align}
To detect the transmitted signal $s[k]$ buried in the reflected RF signal $r[k]$, the JRC transceiver performs the matched filtering. Let $R_s$ denotes the output of the matched filter. Based on Neyman-Pearson criterion, the transceiver decides a detection if $R_s$ exceeds a threshold $\gamma$
    \begin{align}
        R_s&=\sum_{k=0}^{N_{\rm p}-1}r[k]s[k]>\gamma,\qquad\qquad\qquad\qquad\\
        \text{where} \text{ $N_{\rm p}$}&\text{ is the number of signal sample per pulse,}\nonumber\\
        \gamma&=\sqrt{\sigma_{\rm w}^2E_{\rm{p}}}Q^{-1}(P_{\rm FA}),\\	
        Q(x)&=\int_{x}^{\infty}\frac{1}{\sqrt{2\pi}}\exp(-\frac{1}{2}t^2)\;dt\quad
	\\\vspace{-0.5cm}\nonumber
	\end{align}
	is the complement of the cumulative distribution function (CDF) of the standard Gaussian distribution, and $Q^{-1}(x)$ denotes the inverse function of $Q(x)$. The characteristic of $P_{\rm D}$ is given in Theorem 1 below. 

    \textit{Theorem 1:} The value of $P_{\rm D}$ is independent from the JRC signal waveform and only depends on $P_{\rm FA}$ and ENR.
    
\textit{Proof: }From \cite[Ch.~4]{kay1998fundamentals}, $P_{\rm D}$ can be expressed by $P_{\rm FA}$ and ENR as
\begin{align}
	P_{\rm D}&=Q\Big[Q^{-1}(P_{\rm FA})-\sqrt{10^{{\rm ENR}/10}}\;\Big].
	\\\vspace{-0.5cm}\nonumber
	\end{align}
    Accordingly, $P_{\rm D}$ only depends on $P_{\rm FA}$ and ENR, and is independent from the JRC signal waveform.\hfill$\blacksquare$
    
	Therefore, varying the sub-pulse frequencies and durations does not affect the detection performance of the sensing function, as long as the signal energy per each pulse is fixed. Note that, however, varying the sub-pulse durations may cause some increase in the peak-to-average power ratio (PAPR), which reduces the efficiency of the power amplifier at the front end of the JRC transceiver and communication receiver. Therefore, the sub-pulse selection set ${\bf{\mathcal{D}}}$ should be carefully designed to avoid an excessive PAPR value.
	\vspace{-0.2cm}
	\section{Data Demodulation at the Communication Receiver}
	\label{sec:Data Demo}
This section presents the detailed procedure of the proposed data demodulation schemes and its advantages, including being unaffected by channel estimation error (i.e., the channel estimation is not required), and robustness against the Doppler shift or timing offset between the transceiver and the communication receiver.
\vspace{-0.2cm}
	\subsection{Time-Frequency Analysis Technique}
		\begin{figure}[t]
		\centering
		\includegraphics[width=0.7\linewidth]{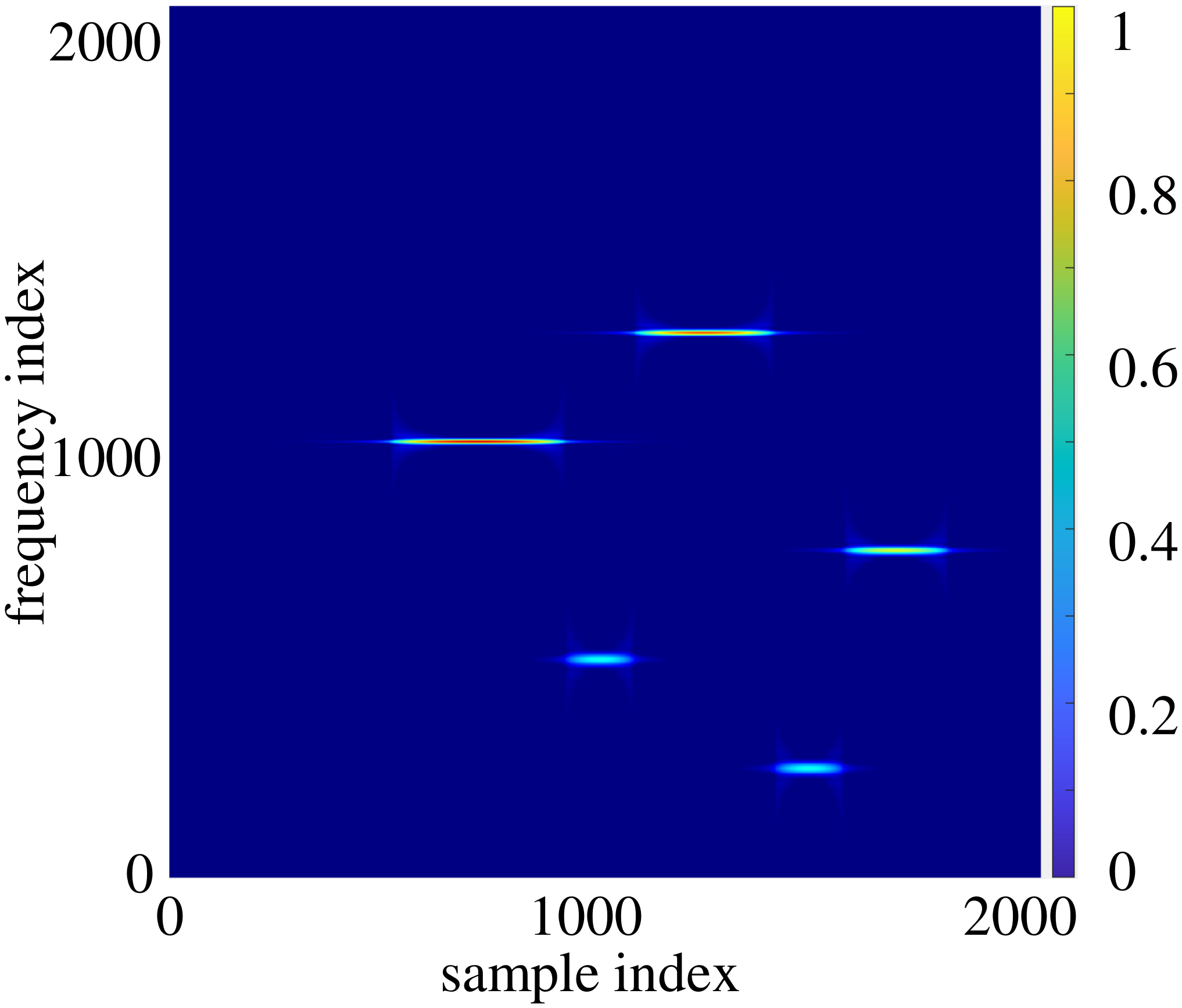}
		\vspace{0.2cm}
		\caption{CWD-TFI of a Costas signal with ${\bf{F}}=[(4,2,5,1,3)\times f_{\rm f}]$, ${\bf{T}}=(5,2,4,2,3)\times 1/(25f_{\rm s})$, $f_{\rm f}=f_{\rm s}/16$, and a starting point of $520$.}
		\label{fig:CWD_TFI}
	\end{figure}
	The CWD is one of the bilinear TFA techniques that is more robust against noise than the linear ones, thanks to the auto-correlation operation in their formulas. Moreover, the CWD does not have the limitation of the cross-term effect that can obscure essential features of the signal. Therefore, in this work, we use the CWD to generate the CWD-TFI, which represents the time-frequency characteristic of the signal. Specifically, the CWD is mathematically given by\cite{choi1989improved}
	\begin{align}
	\label{CWD_1}
	\text{CW}[k,n] & \!=\! 2\sum_{\tau=-\infty}^{\infty}W_{N}(\tau)e^{-j2\pi n\tau/N}\qquad\qquad\qquad\qquad\qquad\nonumber\\
	&\times\Bigg[\sum_{\mu=-\infty}^{\infty}W_{M}(\mu)\frac{1}{\sqrt{4\pi\tau^{2}/\sigma}}e^{-\frac{\mu^{2}}{4\tau^{2}/\sigma}}\nonumber\\
	&\times r[k+\mu+\tau]r^{*}[k+\mu-\tau] \Bigg],
	\end{align}
	where $k$ and $n$ are the time index and frequency index, respectively, $W_{N}(\tau)$ is a symmetrical window function with non-zero values for the range of $-N/2\leq\tau\leq N/2$, while $W_{M}(\mu)$ is a rectangular window function with a value of $1$ for the range of $-M/2\leq\mu\leq M/2$, and $\sigma$ is the scaling factor.
	Similar to \cite{hoang2019automatic}, the value of $N$ and $M$ are often chosen as a power of $2$ to minimize the computational cost (by exploiting the fast Fourier transform algorithm). Furthermore, $\sigma=1$ is used to achieve a balance between frequency resolution and the cross-terms reduction in the CWD.
	\vspace{-0.2cm}
	\subsection{CWD-TFI Preprocessing}
	To generate the CWD-TFI, similar to \cite{hoang2019automatic}, we capture $N_{\rm s}=2048$ consecutive signal samples, and assume these $N_{\rm s}$ samples contain a complete signal pulse, and the signal pulse starts from an arbitrary point within the $N_{\rm s}$ samples. Fig. \ref{fig:CWD_TFI} illustrates the CWD-TFI of a Costas signal with a hopping sequence of ${\bf{F}}=[(4,2,5,1,3)\times f_{\rm f}]$, a sub-pulse duration sequence of ${\bf{T}}=(5,2,4,2,3)\times 1/(25f_{\rm s})$, a fundamental frequency of $f_{\rm f}=f_{\rm s}/16$, and a starting point of $520$. For brevity, the signal uses a Costas array to generate the FH sequence is referred to as a Costas signal. As can be seen, the CWD-TFI clearly demonstrates the time-frequency characteristic of the Costas signal by horizontal lines of different lengths and vertical positions. 
	
\begin{figure}[t]
		\centering
		\includegraphics[width=0.9\linewidth]{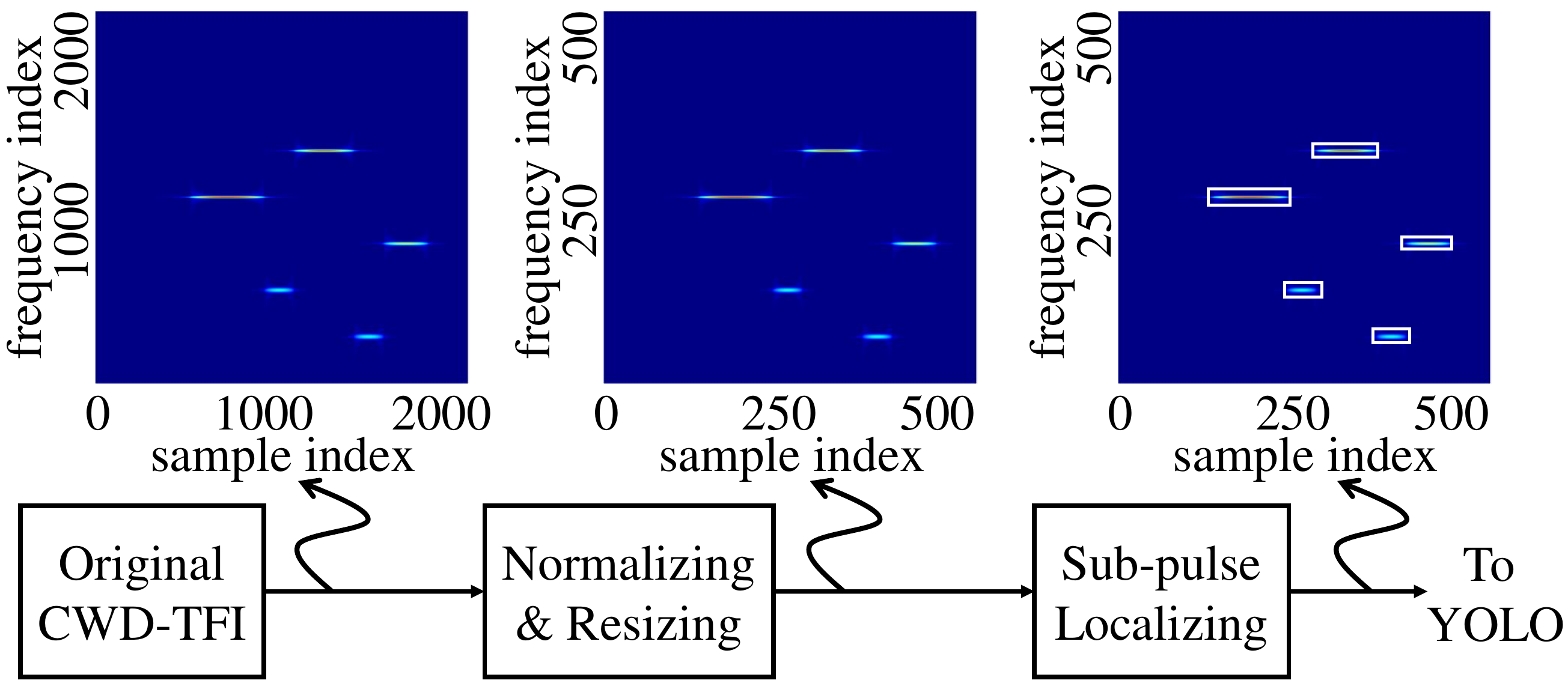}
		\vspace{0.2cm}
		\caption{CWD-TFI preprocessing procedure for\\the Costas signal in Fig. \ref{fig:CWD_TFI}.}
		\label{fig:img_resize}
	\end{figure}

	Fig. \ref{fig:img_resize} shows the CWD-TFI processing procedure used to generate the input to the YOLO detection system. First, using the nearest neighbor interpolation, the CWD-TFI of $N_{\rm s}\times N_{\rm s}$ pixels is resized to $L\times L$ pixels, which is the input size of the YOLO detection system, where $L=500$ pixels is the width of the resized CWD-TFIs. Then, the signal sub-pulse locations in the CWD-TFI are specified by the bounding boxes (i.e., white-color rectangles in Fig. \ref{fig:img_resize}), as a requirement for the training of the YOLO. Note that the localization of the bounding boxes is not needed in the CWD-TFI processing procedure in the testing phase, because the bounding boxes are parts of the output of the trained YOLO detection system.
	\vspace{-0.3cm}
	\subsection{``You only look once'' (YOLO) Object Detection System}
 Among the object detection algorithm for determining the objects types and their locations in the image, YOLO \cite{redmon2016you} is well-known for its real-time detection speed, which is critical in tactical applications, such as military JRC systems. The real-time detection speed is achieved by splitting the input image into a grid of cells and predicting an object and its bounding box on each cell, instead of performing region proposal as in other object detection systems. Details of the YOLO detection system can be found in \cite{redmon2016you}.
 
\begin{figure}[t]
		\centering
		\includegraphics[width=0.65\linewidth]{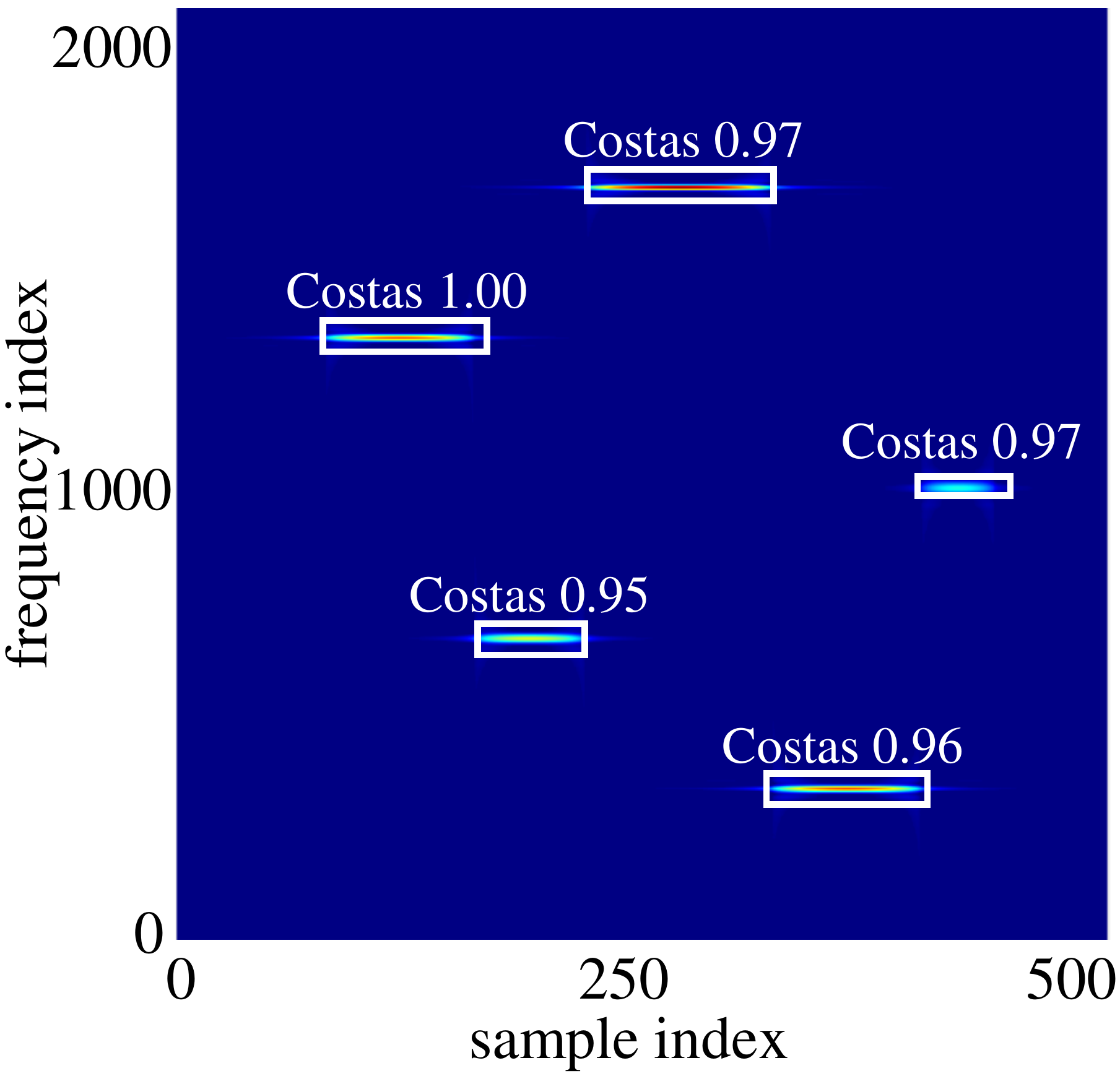}
		\vspace{0.2cm}
		\caption{YOLO's output for a Costas signal with ${\bf{F}}=[(4,2,5,1,3)\times f_{\rm f}]$, ${\bf{T}}=(4, 3, 5, 4, 2)\times 1/(25f_{\rm s})$, $f_{\rm f}=f_{\rm s}/16$, and a starting point of $259$.}
		\label{fig:Yolo_result}
	\end{figure}

 Fig. \ref{fig:Yolo_result} shows the YOLO's output for the CWD-TFI of a Costas signal with an FH sequence of ${\bf{F}}=[(4,2,5,1,3)\times f_{\rm f}]$, a sub-pulse duration sequence of ${\bf{T}}=(4, 3, 5, 4, 2)\times 1/(25f_{\rm s})$, a fundamental frequency of $f_{\rm f}=f_{\rm s}/16$, and a starting point of $259$. As can be seen, the output contains not only the signal type (i.e., ``Costas'') but also the locations of the signal objects (i.e., the white-color rectangular bounding boxes) and the probabilities of the detection.
\subsection{Data Demodulation Technique}
\label{subSec:Demodulation}
As aforementioned, the output from the YOLO detection system contains the bounding boxes, which determine the position, horizontal, and vertical width of the signal objects in the CWD-TFI. The $i$th left-most bounding box, corresponds to the $i$th sub-pulse, is localized by a tuple of $(x^{\rm min}_{i}, x^{\rm max}_{i},y^{\rm min}_{i}, y^{\rm max}_{i})$. Recall that for the CWD-TFI, $L$ pixels in the vertical axis correspond to a frequency range from $0$ to $f_{\rm s}/2$, while $L$ pixels in the horizontal axis correspond to a time duration of $N_{\rm s}\times(1/fs)$, the frequency and duration of the $i$th sub-pulse can be calculated by
	\begin{align}
	    f_i&=\frac{f_{\rm s}(y^{\rm min}_{i}+y^{\rm max}_{i})}{4L},\nonumber\\
	    \triangle t_{i}&=\frac{N_{\rm s}(x^{\rm max}_{i}-x^{\rm min}_{i})}{Lf_{\rm s}}.
	    \\\vspace{-0.5cm}\nonumber
	\end{align}
	By comparing the combination of $f_i$ and $\triangle t_{i}$ values with the selections in the codebook $\mathcal{S}$ or $\mathcal{S}_{\rm C}$, the embedded symbol in each signal pulse can be determined. 
	
	It is worth noting that this demodulation technique does not require any knowledge about the channel between the JRC transceiver and the communication receiver, and hence the channel estimation is not necessary. Moreover, since our data demodulation technique is based on the CWD-TFI, its performance is not affected by the Doppler shift or timing offset between the transceiver and the communication receiver. This is because the Doppler shift or timing offset merely creates vertical or horizontal shifts of the whole set of lines, respectively. On the other hand, the shapes of the lines in the CWD-TFI remain unchanged.
	\vspace{-0.3cm}
	\section{Performance Evaluation}
	\label{Sec:Simulation}
\begin{table}[t]
		\small	
		\caption{\textsc{Parameters values.}}
		\vspace{0.2 cm}
		\begin{tabular}{|P{1.3cm}|P{3.2cm}|P{3.1cm}|}
			\hline 
			\textbf{Parameter} & \textbf{Description}& \textbf{Value}\\ \hline
			\break$N_{\rm f}$&Number of sub-pulses per pulse&\break 5\\
			\hline
			$N_{\rm C}$&Number of Costas array&40\\
			\hline
			\break${\bf{\mathcal{D}}}$&Sub-pulse duration selection set& $(1,1.5,2,2.5,3)\times 1/(25f_{\rm s})$\\
			\hline
			$f_{\rm f}$&Fundamental frequency&$f_{\rm s}/16$\\
			\hline
			\break$L$&Resized CWD-TFI width (pixels)&\break500\\
			\hline 
		\end{tabular}
		\label{Tab:notation}
\end{table}
Table. \ref{Tab:notation} shows the parameter values used to produce the simulation results. Note that the number of Costas array $N_{\rm C}$ is not a freely set value, but depends on $N_{\rm f}$, the number of sub-pulses per pulse. For instance, $N_{\rm C}=40$ for $N_{\rm f}=5$. The list of available Costas arrays and their construction can be found in \cite[Ch.~5]{levanon2004radar}. To train the YOLO detection system, for each data embedding scheme, we generate $2700$ signals with SNR ranging from $-6$ dB to $10$ dB and a step size of $2$ dB. The $2700$ signals are divided into a training set of $2160$ signals ($80\%$ of the total) and a validation set of $540$ signals (20\% of the total). For the testing data, we generate $3300$ signals for each embedding scheme, with SNR ranging from $-10$ dB to $10$ dB and a step size of $2$ dB.
\begin{figure}[t]
		\centering
		\includegraphics[width=0.9\linewidth]{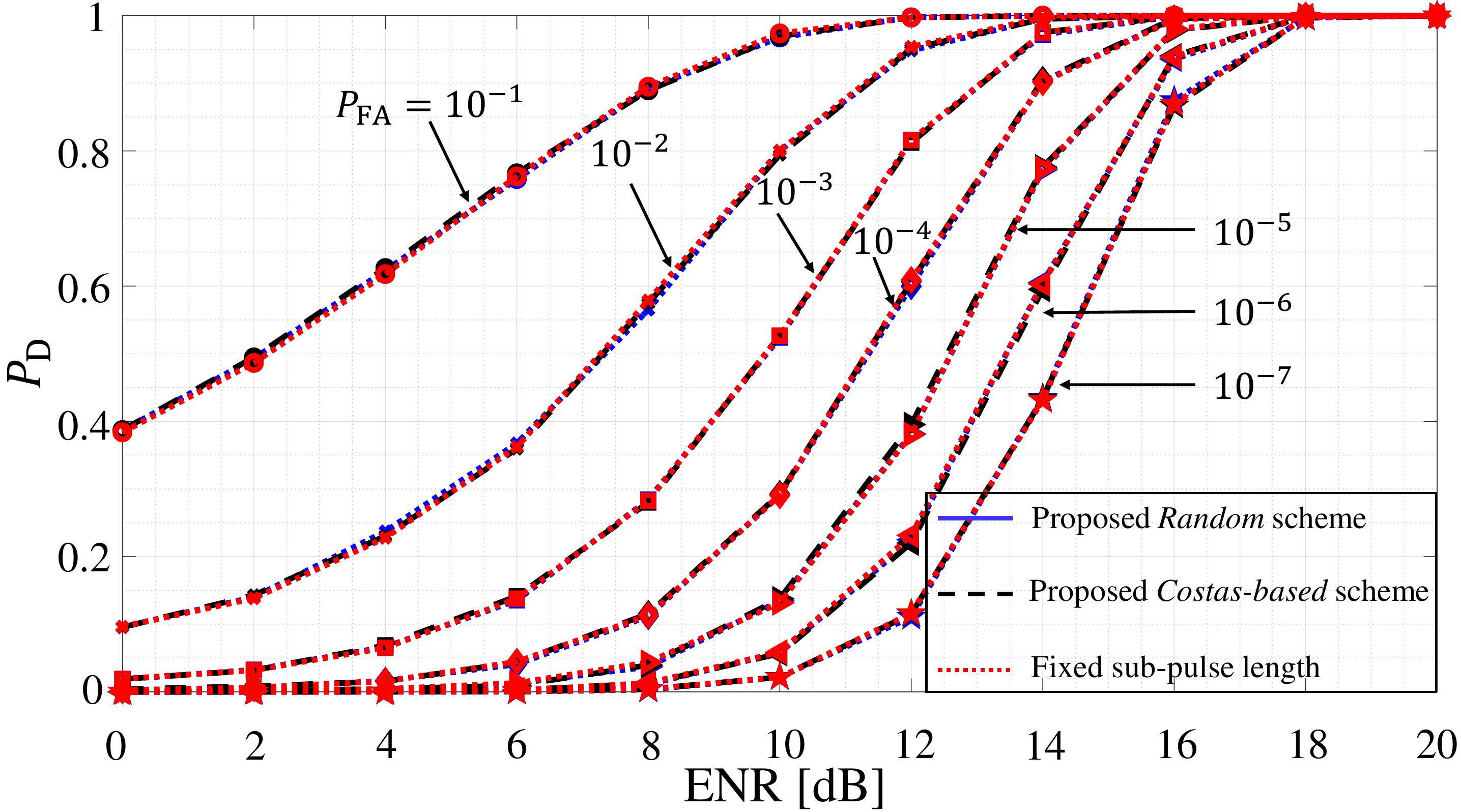}
		\vspace{0.2cm}
		\caption{Detection probability of the JRC sensing function for different signal waveforms, $P_{\rm FA}$, and ENR values.}
		\label{fig:PD}
	\end{figure}
\begin{figure}[t]
		\centering
		\includegraphics[width=0.9\linewidth]{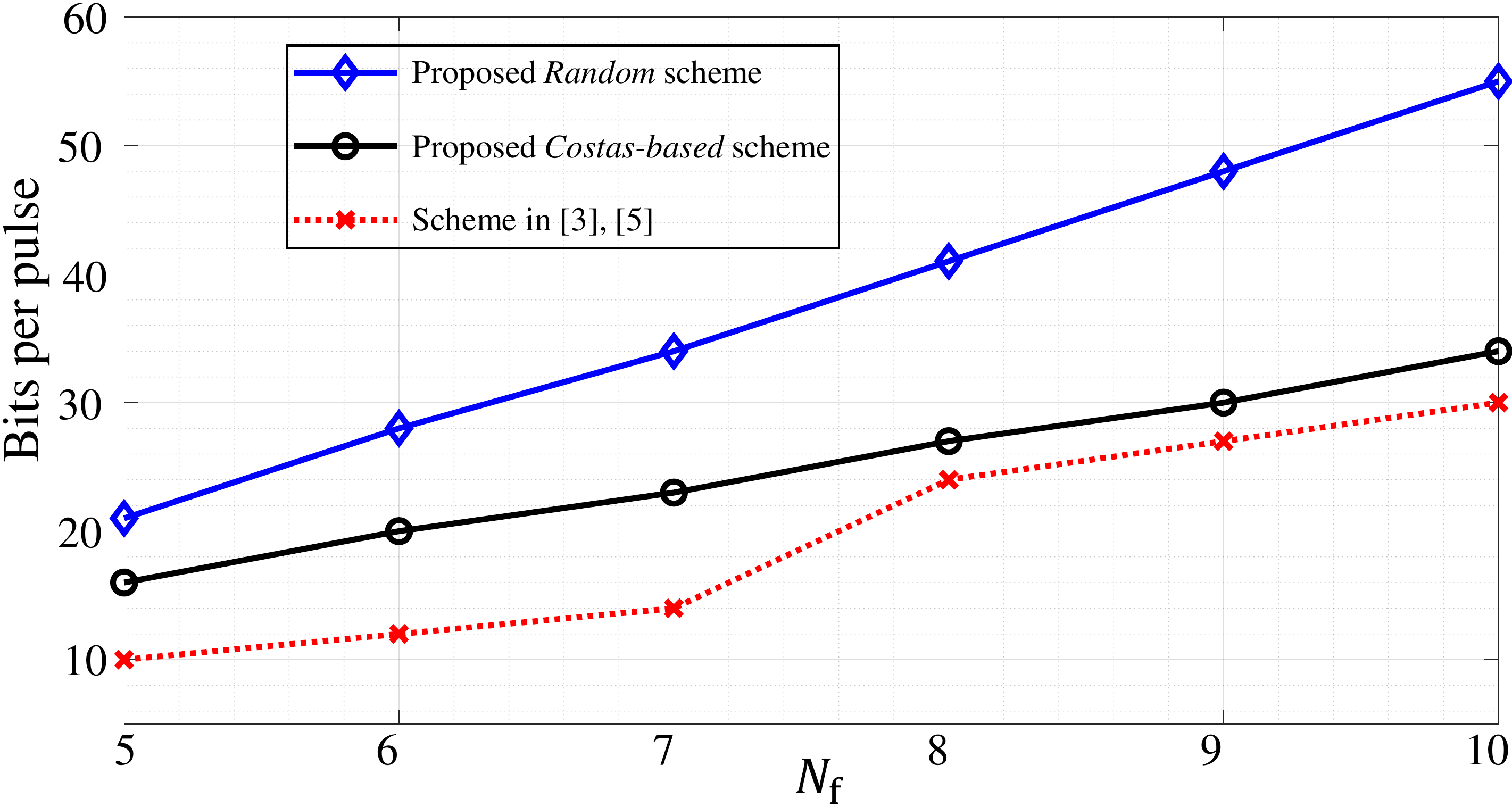}
		\vspace{0.2cm}
		\caption{Maximum number of bits per pulse for different\\
data embedding schemes.}
		\label{fig:Rate}
	\end{figure}	

Fig. \ref{fig:PD} shows the detection probability of the JRC sensing function for different signal waveforms (i.e., \textit{Random} scheme, \textit{Costas-based} scheme, and fixed sub-pulse length), $P_{\rm FA}$, and ENR values. The results are obtained by averaging $10,000$ independent trials. As can be seen, for each $P_{\rm FA}$ value, the three curves almost overlap, meaning the detection probability only depends on $P_{\rm FA}$ and ENR but not the signal waveform. Therefore, varying the sub-pulse durations does not affect the detection capability of the sensing function of the FH RJC system, as long as the signal energy per each pulse is fixed.

Fig. \ref{fig:Rate} demonstrates the maximum number of bits that can be transmitted over a signal pulse of the two proposed embedding schemes and those in \cite{huang2020majorcom,baxter2018dual}. As can be seen, the \textit{Random} scheme can embed a significantly higher amount of data than the \textit{Costas-based} scheme, while the \textit{Costas-based} scheme can convey a slightly higher number of data bits than the scheme in \cite{huang2020majorcom,baxter2018dual}. That is because the schemes in this paper use both sub-pulse durations and frequencies to embed signals. On the other hand, the technique in \cite{huang2020majorcom,baxter2018dual} focuses on using antenna permutations for data embedding. Therefore, the proposed techniques are more suitable when the number of antennas is limited, while the technique in \cite{huang2020majorcom,baxter2018dual} is suitable in the case of an extensive number of antennas is available.
    \begin{figure}[t]
		\centering
		\includegraphics[width=0.9\linewidth]{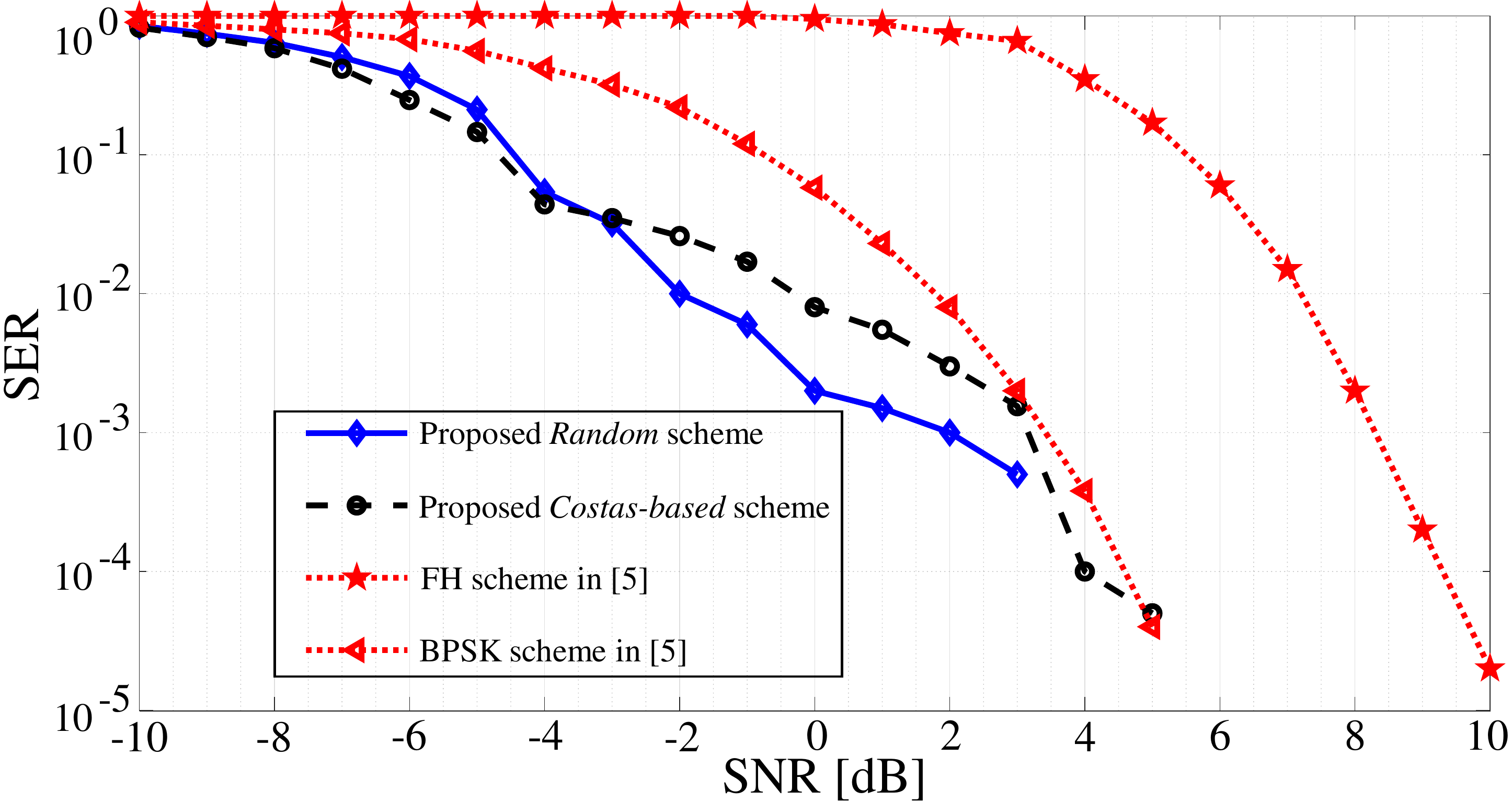}
		\vspace{0.2cm}
		\caption{Symbol error rate (SER) of different embedding\\and demodulation schemes.}
		\label{fig:SER}
	\end{figure}
	
Fig. \ref{fig:SER} illustrates the symbol error rate (SER) of the two proposed schemes and those of the techniques in \cite{baxter2018dual} as functions of the signal-to-noise ratio (SNR). As can be seen, the two proposed schemes have lower SERs compared to those of the techniques in \cite{baxter2018dual} for all SNRs. Distinctively, at low SNRs (i.e., $\leq-2$ dB), the two proposed schemes have similar SERs, and are about $4$ dB and $10$ dB better than those of the FH and BPSK techniques in \cite{baxter2018dual}, respectively. This is because the proposed techniques use the CWD, which contains the auto-correlation operation, to generate CWD-TFIs of the signals. As such, the proposed techniques are more robust against noise compared to the other techniques in \cite{baxter2018dual}.
	\section{Conclusions}
We have proposed novel approaches for embedding and demodulating information bits in FH JRC systems. For data embedding, we have used both sub-pulse durations and frequencies to increase the data transmission rate. For data demodulation, to reduce the error rate, we have proposed a CWD-TFI and YOLO-based demodulation scheme that does not require channel estimation and is robust against Doppler shift and timing offset between the JRC transceiver and communication receiver. Simulation results have shown that our proposed techniques achieve higher data transmission and lower symbol error rates than the existing ones. 

\end{document}